\documentclass[12pt]{article}
\newcommand{\bfr}{{\bf r}}

\newcommand{\bd}{\begin{displaymath}}
\newcommand{\ed}{\end{displaymath}}
\newcommand{\be}{\begin{equation}}
\newcommand{\ee}{\end{equation}}
\newcommand{\bq}{\begin{quote}}
\newcommand{\eq}{\end{quote}}
\newcommand{\ben}{\begin{enumerate}}
\newcommand{\een}{\end{enumerate}}
\newcommand{\bi}{\begin{itemize}}
\newcommand{\ei}{\end{itemize}}
\newcommand{\bdes}{\begin{description}}
\newcommand{\edes}{\end{description}}

\begin{document}
\setlength{\baselineskip}{16.5pt}
\title{Reflections on the Spatiotemporal Aspects\\
of the Quantum World\footnote{Invited talk at the First Meeting on the 
Interface of Gravitational and Quantum 
Realms held at IUCAA, Pune (India), December 17--21, 2001. 
To appear in Mod. Phys. Lett. A.}}
\author{Ulrich Mohrhoff\\
Sri Aurobindo International Centre of Education\\
Pondicherry-605002 India\\
\normalsize\tt ujm@satyam.net.in}
\date{}
\maketitle
\begin{abstract}
\noindent The proper resolution of the so-called measurement problem requires a ``top-down'' 
conception of the quantum world that is opposed to the usual ``bottom-up'' conception, 
which builds on an intrinsically and maximally differentiated manifold. The key to 
that problem is that the fuzziness of a variable can manifest itself only to the extent that 
less fuzzy variables exist. Inasmuch as there is nothing less fuzzy than the metric, this 
argues against a quantum-gravity phenomenology and suggests that a quantum theory 
of gravity is something of a contradiction in terms---a theory that would make it possible 
to investigate the physics on scales that do not exist, or to study the physical 
consequences of a fuzziness that has no physical consequences, other than providing a 
natural cutoff for the quantum field theories of particle physics.
\setlength{\baselineskip}{14pt}
\end{abstract}

\section{\large INTRODUCTION}
\label{Intro}
At present the almost single constraint on speculations about the interface of the gravitational 
and quantum realms is consistency with general relativity (GR) and the standard model 
(SM) as effective theories for their respective realms. The {\it meaning\/} of quantum 
mechanics (QM) is not generally considered a constraint. The prevalent attitude seems 
to be that any interpretation will do as long as it permits asking the pertinent questions, 
or that the final resolution of interpretational issues must wait for a unified quantum 
theory of all forces including gravity. I shall argue instead that the proper resolution of 
the so-called measurement problem requires a ``top-down'' conception of the quantum 
world that is opposed to the usual ``bottom-up'' conception, which builds on an 
intrinsically and maximally differentiated manifold. The key to that problem is that the 
fuzziness of a variable can manifest itself only to the extent that less fuzzy variables exist. 
Inasmuch as there is nothing less fuzzy than the metric, this argues against a 
quantum-gravity phenomenology and suggests that a quantum theory of gravity is 
something of a contradiction in terms---a theory that would make it possible to 
investigate the physics on scales that do not exist, or to study the physical consequences 
of a fuzziness that has no physical consequences, other than providing a natural cutoff for the 
quantum field theories of particle physics.

Section 2 ties together the probabilistic nature of quantum ``states,'' the positional 
fuzziness that ``fluffs out'' matter, and the extrinsic nature of quantum variables. Section~3 
requires nothing more sophisticated than a two-slit interference experiment to show that 
space cannot be a self-existent and intrinsically partitioned expanse, and that the 
intrinsically differentiated spatiotemporal background of classical field theory cannot be part of a 
description of the spatiotemporal aspects of the quantum world. Section~4 disposes of a vicious 
regress that appears to arise from the extrinsic nature of quantum variables, and establishes the 
existence of a macroworld---a system of causally connected properties that are effectively 
detached from the facts by which they are indicated. It further demonstrates that the quantum 
world is only finitely differentiated both spacewise and timewise, and that as a consequence it 
ought to be regarded as constructed from the top down, by the spatiotemporal differentiation of 
an intrinsically undifferentiated reality, rather than from the bottom up, on an intrinsically and 
maximally differentiated manifold.

Section~5 tries to bring our intuitions in line with the spatiotemporal aspects of the 
quantum world by stressing that the so-called ``point particles'' are {\it formless\/} 
objects, and that space exists {\it between\/} them---it is spanned by their (more or less 
fuzzy) relations. Section~6 deals with the physical roots of the metric properties of the 
quantum world and the meaning of ``second quantization.'' Section~7 discusses how the 
quantum world relates to its substance. Section~8 argues against the need for a quantum 
theory of gravity, and Sec.~9 reflects on the possible meaning of ``the wave function of 
the (early) universe.''

\section{\large INDEFINITENESS}
\label{Indef}
Let me begin by denouncing a didactically disastrous approach to QM. This starts with 
the observation that in classical physics the state of a system is represented by a point 
$\cal P$ in some phase space, and that the system's possessed properties are represented 
by the subsets containing~$\cal P$. Next comes the question, what are the 
quantum-mechanical counterparts to $\cal P$ and the subsets containing~$\cal P$ {\it 
qua representations of an actual state of affairs and possessed properties\/}, respectively? 
Once we accept this as a valid question, we are on a wild-goose chase.

If at all we need to proceed from classical physics, the right way to do so is to point out 
that every classical system is associated with a probability measure, that this is 
represented by a point~$\cal P$ in some phase space, that observable properties are 
represented by subsets, and that the probability of observing a property is~1 if the 
corresponding subset contains~$\cal P$; otherwise it is~0. Next comes the question, what 
are the quantum-mechanical counterparts to $\cal P$ and the subsets containing~$\cal 
P$ {\it qua representations of a probability measure and observable properties\/}, 
respectively? Once we have the answer we are ready for the next question: Is it 
possible to reinterpret the quantum-mechanical counterpart to~$\cal P$ as representing 
an actual state of affairs connoting a set of possessed properties? Because the classical 
probability measure assigns trivial probabilities (either 0 or~1), it is possible to think of 
it as an actual state of affairs. Because quantal probability measures generally assign 
nontrivial probabilities, it is not possible to similarly reinterpret the quantum 
counterpart to~$\cal P$.

Whence the nontrivial probabilities? One of the most obvious features of the world is the 
stability of matter, by which I mean the existence of spatially extended material objects 
that neither explode nor implode the moment they are formed. We know that the world 
owes this feature to the indefiniteness of its relative positions. Together with the 
exclusion principle it ``fluffs out'' matter~\cite{Lieb}. The proper way of dealing with 
indefinite values is to make counterfactual probability 
assignments~\cite{Mohrhoff00,Mohrhoff01}. If we say that a variable has an ``indefinite 
value,'' what we mean is that it does not have a value (inasmuch as no value is indicated) 
but that it {\it would\/} have a value if one {\it were\/} indicated, and that positive 
probabilities are associated with at least two possible values. While the reference to 
counterfactuality cannot be eliminated, it may be shifted from values that are only 
counterfactually indicated to values that are only counterfactually indefinite: If a certain 
measurement is performed on an ensemble of identically ``prepared'' systems and the 
results exhibit a positive dispersion, the value of the measured variable would be 
indefinite for each system if the measurement were not performed.

The occurrence of irreducible probabilities in a fundamental physical theory thus is a 
direct consequence of the positional indefiniteness to which matter owes its stability. 
So is the {\it extrinsic\/} nature of values. If a variable sometimes does and sometimes does not 
have a value, a criterion is called for, and this is the existence of a value-indicating fact. Such a 
variable possesses a value only if, when, and to the extent that, a value is indicated (by an 
actual event or state of affairs). A fundamental physical theory that is essentially a probability 
algorithm presupposes actual events (or states of affairs) not once but twice: as events to which 
probabilities are assigned, and as events on the basis of which probabilities are assigned.

What is a fact? Dictionaries define facts in epistemological terms. The {\it Concise Oxford 
Dictionary} (8th edition, 1990), for instance, defines ``fact'' as a thing that is {\it known\/} to 
have occurred, to exist, or to be true; a datum of {\it experience\/}; an item of verified {\it 
information\/}; a piece of {\it evidence\/}. Are the editors of dictionaries intent on convincing us 
that the existence of facts presupposes knowledge or experience? To find out, let's look up these 
terms: ``Experience'' is the ``actual observation of or practical acquaintance with {\it facts\/} 
or events''. ``Knowledge'' is ``awareness or familiarity gained by experience (of a person, {\it 
fact}, or thing)'' (italics supplied). These definitions are obviously circular, which might suggest a 
{\it mutual\/} dependence: All knowledge is knowledge of facts, and all facts are known facts. 
This may well be so, but it is beside the point.

Physics is concerned with laws, and hence with nomologically possible worlds (that is, with worlds 
consistent with the laws). It does {\it not\/} tell us which of these possible worlds corresponds to 
the actual world. In classical physics the actual world can be identified by initial or final 
conditions (or both), which we must gather from observational data. In quantum physics the 
identification of the actual world requires the identification of (i)~the actual property-indicating 
facts (a.k.a. ``measurements'') and (ii)~the property indicated by each such fact (a.k.a. the 
``measurement result''). These too we need to gather from observations. Does this imply that 
quantum physics presupposes conscious observers? If the answer is negative for classical 
physics, it is equally negative for quantum physics.

The point is that a physical theory cannot identify the actual world, cannot characterize it, 
cannot distinguish it from the merely possible worlds, and thus cannot define actuality, reality, 
existence, or factuality. The existence of an actual world is the ultimate given. But what matters 
is {\it that\/} it is given, not for or to whom it is given, nor whether there is anyone to whom it 
is given. No theory can explain why there is anything at all, rather than nothing. This is why QM 
cannot explain the property-indicating facts, which it takes for granted.

That QM presupposes the existence of such facts is readily seen: The probability that a variable 
$Q$ has the value $v$ is the product of two probabilities---the probability that any one of the 
possible values of $Q$ is indicated, and the probability that the indicated value is~$v$, given 
that a value is indicated. QM is exclusively concerned with probabilities of the latter type. It does 
not assign a probability to the occurrence of a value-indicating event, let alone specify sufficient 
conditions for such an occurrence. If QM is a fundamental and universal theoretical framework, 
this means that the value-indicating events presupposed by QM are {\it uncaused\/}. Any 
attempt to ``explain why events occur''~\cite{Pearle79} is therefore a waste of time.

What needs to be shown instead is that QM is consistent: There are possible events to which 
actuality can be attributed consistently. In other words, there are properties that can be 
consistently regarded as intrinsic, and hence as capable of indicating extrinsic properties. The 
question is not, ``how it is that probabilities become facts''~\cite{Treiman} but, which 
properties of which objects should be considered factual {\it per se\/}. This question is 
addressed in Sec.~4.

\section{\large POSITIONS ARE PROPERTIES, NOT SUBSTANCES}
\label{PPNS}
Making sense of QM is not so much a question about the ontological status of density 
operators---they are just sophisticated probability measures---as a question about the 
ontological status of {\it the space and time coordinates\/} that appear as arguments of 
density operators in the position representation. Classical field theory has instilled in us 
the disastrous habit of thinking of anything that depends on positions and times as 
something that exists at those positions and times, which implies the independent 
existence of those positions and times. Yet it should be obvious that the probability for 
something to happen in a given region of space at a given time is not something that 
exists in that region or at that time, and it stands to reason that the same is true of the 
algorithm that permits us to calculate this probability. Can we nevertheless postulate the 
independent existence of the intrinsically differentiated spatiotemporal background of 
classical field theory?

Showing that the answer is negative takes nothing more elaborate than a two-slit experiment 
with electrons~\cite{Feynmanetal65}. No electron is detected in the absence of the electron 
source in front of the slit plate, and no electron is detected behind the slit plate whenever the 
two slits are closed. This warrants the inference that each detected electron went through 
$L\&R$, the regions defined by the slits considered as one region. At the same time the 
existence of interference fringes demonstrates that each electron went through $L\&R$ 
without going through a particular slit (and, of course, without having been split into parts that 
went through different slits). But if space were something that existed by itself, independently of 
its ``material content,'' and if it were made up of distinct, separate regions, no object could 
have a fuzzy position---a position that is counterfactually and probabilistically distributed over 
disjoint regions. The truth of the proposition ``The electron went 
through~$L\&R$''---symbolically, $e{\rightarrow}L\&R$---would imply the truth of either 
$e{\rightarrow}L$ or $e{\rightarrow}R$.

Interference fringes have been observed using C$_{60}$ molecules and a grating with 
50-nm-wide slits and a 100-nm period~\cite{Arndtetal}. Do we need any further proof 
that $L$ and $R$ cannot be distinct, self-existent ``parts of space,'' and that, 
consequently, space cannot be a self-existent and intrinsically partitioned expanse? 
The proposition $e{\rightarrow}L\&R$ can be true while both $e{\rightarrow}L$ and 
$e{\rightarrow}R$ lack truth values. In other words, a definite relationship---``inside'' or 
``outside''---can exist between the electron's position (qua observable) and the region $L\&R$ 
while no such relationship exists between the electron's position and either 
$L$ or~$R$. In yet other words, $L\&R$ can be real for the electron while neither $L$ nor~$R$ 
(nor the distinction we make between $L$ and~$R$) is real for it.

Although we readily agree that red, or a smile, cannot exist without a red object or a 
smiling face, we just as readily believe that positions can exist without being properties 
of material objects. We are prepared to think of material objects as substances, and we 
are not prepared to think of their properties as substances---except for their positions. 
(A substance is anything that can exist without being the property of something else.) 
There are reasons for these disparate attitudes, but they are psychological and 
neurobiological~\cite{BCCP,QMCCP}. They concern the co-production, by the mind 
and the brain, of the phenomenal world---the world as we perceive it. They do not apply 
to the quantum world, but they certainly make it hard to make sense of it.
As long as we treat $L$ and $R$ as substances that make up $L\&R$, we cannot comprehend 
the behavior of electrons in two-slit experiments. This behavior makes sense only if we treat 
each region of space as a property---the property of being in that region. Then a region exists iff 
it is possessed by at least one object, and the distinction between a region $V$ and its 
complement is real for an object $O$ iff the proposition ``$O$~is in~$V$'' has a truth value 
(that is, if a truth value is indicated). If neither $L$ nor $R$ exists for the electron, nothing can 
``compel'' the electron to go through either $L$ or~$R$.

\section{\large THE MACROWORLD}
\label{MW}
There are objects whose indicated positions are so correlated that each of them is 
consistent with every prediction that is based on previous indicated positions and a 
classical law of motion {\it except\/} when it serves to indicate an unpredictable value. Such 
objects deserve to be labeled ``macroscopic.'' (Note that this definition does not require 
that the probability of finding a macroscopic object where classically it could not be, is 
strictly~0. What it requires is that there be no position-indicating fact that is inconsistent with 
predictions based on a classical law of motion and earlier position-indicating facts.) Since 
between those times at which macroscopic objects serve as pointers their indicated positions are 
predictably correlated, these positions can be considered intrinsic, or factual {\it per se\/}. And 
since before and after each value-indicating transition the position of a pointer can be considered 
intrinsic, factuality can also be attributed to the transition itself. These claims will be 
substantiated in the present section.

The departure of an object $O$ from a classical trajectory can be indicated only if there 
are detectors whose position probability distributions are narrower than~$O$'s. Such 
detectors do not exist for all objects. Some objects have the sharpest positions in 
existence. For these objects the probability of a position-indicating event that is 
inconsistent with a classical trajectory is necessarily very low. It is therefore certain that 
{\it among\/} these objects there will be macroscopic ones.

Since no object has an exact position, it might be argued that even for a macroscopic 
object~$M$ there always exists a small enough region~$V$ such that the proposition 
$M{\rightarrow}V$ lacks a truth value. But this is an error. Among the objects that have the 
sharpest positions in existence there are macroscopic objects, and they are macroscopic (in the 
sense defined above: their nonindicating positions are predictably correlated) because there isn't 
any object that has a (significantly) sharper position. Hence there isn't any object for which $V$ 
is real. But a region exists only if it is real for at least one object. It follows that there exists no 
region~$V$ such that the proposition $M{\rightarrow}V$ lacks a truth value. Such a region may 
exist in our imagination, but it does not exist in the real world.

Now recall why positions are extrinsic: The proposition $O{\rightarrow}V$ may or may 
not have a truth value. One therefore needs a criterion for the existence of a truth value---a 
truth value must be indicated. But one doesn't need a criterion for the existence of a 
truth value if for every {\it existing\/} region~$V$ the proposition $M{\rightarrow}V$ 
{\it has\/} a truth value. Since macroscopic objects satisfy this condition, their positions 
can be consistently considered intrinsic. Since every existing region has a trivial probability of 
containing~$M$, we can make the transition from a probability measure to an actual state of 
affairs, exactly as in classical physics (Sec.~\ref{Indef}). We can think of the positions 
of macroscopic objects (macroscopic positions, for short) as forming a system of causally 
connected properties that are effectively detached from the facts by which they are indicated. 
We can think of this system as a self-contained causal nexus interspersed with transitions (of 
value-indicating positions) that are causally linked to the future but not to the past. And it is to 
this system---and this system alone---that an independent reality can be attributed. Everything 
else exists because it is indicated by the goings-on within this system, for without it no 
indicatable properties exist. (The function of a detector is not only that of indicating a 
position but also that of making real an indicatable position---the detector's sensitive region---by 
possessing it as an intrinsic property---a predictably evolving shape.)

Macroscopic positions are so abundantly and so sharply indicated that they are only 
counterfactually fuzzy. Their fuzziness never evinces itself, through uncaused transitions 
or in any other manner. It exists solely in relation to an imaginary spatial background 
that is more differentiated than the real world. The space over which the position of a 
macroscopic object is ``smeared out'' is never probed. This space is undifferentiated; it 
contains no smaller regions. We may imagine smaller regions, but they have no 
counterparts in the real world. The distinctions we make between them are distinctions 
that nature does not make.

It follows that the quantum world is only finitely differentiated spacewise, and that we 
ought to regard it as constructed {\it from the top down\/}, by a finite process of 
differentiation, rather than from the bottom up, on an intrinsically and maximally 
differentiated manifold. And much the same applies to the world's temporal aspect. 
Time, as everyone knows, is not an independent observable. Time has to be read off of 
deterministically evolving positions---the positions of macroscopic clocks. If these bear a 
residual fuzziness, so do all indicated times. The upshot: The quantum world is maximally 
differentiated neither spacewise nor timewise.

\section{\large SPACE AND THE QUANTUM WORLD}
\label{SQW}
Quarks and leptons are often described as ``pointlike.'' It is important to be clear about 
what this means. It expresses the fact that such an object lacks internal structure. 
Nothing in the formalism of QM refers to the shape of an object that lacks internal 
structure, and the empirical data cannot possibly do so. All that experiments can reveal 
in this regard is the absence of evidence of internal structure. The idea that a 
so-called ``point particle'' is an object that not only lacks internal relations but also has 
the shape of a point thus is unwarranted both theoretically and experimentally. It is, 
besides, seriously misleading, inasmuch as the image of a pointlike object suggests the 
existence of an infinitesimal neighborhood in an intrinsically and maximally 
differentiated manifold. To bring our intuitions in line with the spatiotemporal aspects of 
the quantum world, we ought to conceive of all so-called ``point particles'' as {\it 
formless\/} objects. What lacks internal relations also lacks a shape.

It follows that the shapes of material objects resolve themselves into sets of (more or less 
fuzzy) spatial relations between formless objects, and that space itself is the totality of 
such relations---relative positions and relative orientations. It further follows 
that the corresponding relata {\it do not exist in space\/}. Space contains, in the proper, 
set-theoretic sense of ``containment,'' the forms of all things that have forms---for forms 
are sets of spatial relations---but it does not contain objects over and above their 
forms; {\it a fortiori\/} it does not contain the formless constituents of matter. Instead, 
space exists {\it between\/} them; it is spanned by their relations.

Apart from being of philosophical interest, these conclusions may be important for a 
better understanding of the quantum/gravity interface, inasmuch as they provide us 
with a way of thinking about the spatial and temporal aspects of the quantum world 
that is consistent with their finite differentiation. The quantum world with its fuzzy 
spatial relations does not ``fit'' into the self-existent and maximally differentiated 
expanse of classical space; the possibility of thinking of the relata as points and 
embedding them in a single manifold exists only for definite (``sharp'') spatial relations. 
A clear distinction should therefore be made between the existing (more or less fuzzy) 
spatial relations that constitute space, and the purely imaginary space that comes with 
each localizable object~$O$ and contains the unpossessed exact positions relative to~$O$. 
These imaginary spaces are delocalized relative to each other: The unpossessed exact 
positions relative to~$O$ are fuzzy relative to any object other than~$O$.

\section{\large THE PHYSICAL ROOTS OF THE METRIC}
\label{PRM}
The formal expression of indefiniteness (fuzziness) through nontrivial probability 
assignments determines the kinematical aspects of QM up to the number 
field~\cite{JustSo}. Together with the conservation of probability for freely propagating 
particles, the kinematics determines the dynamical framework of QM {\it including\/} 
the number field. Here is (in outline) how the latter may be shown:

Consider a series of measurements. We assign probabilities to the possible results of the 
final measurement in conformity with the probability algorithm that constitutes the 
kinematics. If the intermediate measurements are not performed but the histories that 
lead to a possible result of the final measurement are defined in terms of the possible 
results of the intermediate measurements, the probability of any possible final result will 
be given by the (absolute) square of a sum over all histories that lead to this result. Each 
history contributes an amplitude, which may be real or complex.

Next consider the limit of a series of unperformed position measurements on an isolated 
scalar particle, in which the histories become continuous trajectories. Suppose $s$ 
parametrizes such a trajectory, and $ds_1$ and $ds_2$ label adjoining infinitesimal 
segments. The probability algorithm requires us to multiply the amplitudes associated 
with successive segments of a history, so that
\bd
A(ds_1+ds_2)=A(ds_1)A(ds_2).
\ed
Hence the amplitude for propagation along an infinitesimal path segment can be written as 
$A(ds)=\exp(z\,ds)$, and the amplitude for propagation along a path $\cal C$ can be 
written as $ A(s[{\cal C}])=\exp(zs[{\cal C}])$. But if $z$ had a real part, the probability 
of finding the particle anywhere would not be conserved; it would either increase or 
decrease exponentially with time. Thus $A(s[{\cal C}])$ must be a phase factor 
$\exp(ibs[\cal C])$.

It should perhaps be stated that no contradiction exists between (i)~concluding that the 
intrinsically differentiated spatiotemporal background of classical field theory does not exist in 
the quantum world and (ii)~introducing parametrized continuous trajectories. For one, each 
trajectory is defined counterfactually as a sequence of results of unperformed (and even 
unperformable) position measurements. For another, the particle trajectories ``exist'' in the 
purely imaginary space of unpossessed exact positions mentioned at the end of the previous 
section. (In this case the origin of that ``space'' is any object relative to which the particle got 
precisely localized by the first measurement.) By summing over all continuous paths leading 
from $(x_1,t_1)$ to $(x_2,t_2)$, we can calculate the propagator $K(x_2,t_2;x_1,t_1)$, which 
determines the probability of detecting at $(x_2,t_2)$ a particle having last been ``seen'' at 
$(x_1,t_1)$~\cite{FH65,Feynman48}. The probability for something to happen at $(x_2,t_2)$, 
recall, is not something that exists at $(x_2,t_2)$, nor is the spacetime location $(x_2,t_2)$ 
something that exists by itself. It exists (for the particle) if something indicates that at the exact 
time~$t_2$ the particle has the exact position~$x_2$. The very fact that in calculating 
$K(x_2,t_2;x_1,t_1)$ we sum over all particle trajectories connecting the two locations, implies 
that the distinctions we make between the trajectories are distinctions that nature does not 
make. If the particle is detected at $(x_1,t_1)$ and $(x_2,t_2)$, and if there isn't any matter of 
fact about its intermediate whereabouts, the particle travels along ``all of them'' without 
traveling along any of them, in the same sense in which the electron goes through $L\&R$ 
without going through either $L$ or~$R$~\cite{Clicks}.

We are now ready to address the physical roots of the metric properties of the quantum 
world. Consider a scalar particle and a particular path. As the particle travels along this path 
(counterfactually), $s$~increases, and $\exp(ibs)$ rotates in the complex plane. Let us say that 
every time this phase factor completes a cycle, the particle ``ticks.'' If the particle is free, it 
singles out a class of uniform time parameters---those for which the number of ticks per second 
is constant. Different particles may tick at different rates, which are related to the standard rate 
of one tick per second by the species-specific constant factor~$b$.

So much for the physical roots of mass and proper time. Our next task is to determine 
the physical origin of the spatial part of the metric. If space is a set of spatial relations 
then there are no absolute positions, nor is there anything like absolute rest. Hence all 
inertial coordinate systems are created equal. It can be shown~\cite{SU} that, as a 
consequence, the proper-time interval $ds$ and inertial coordinates are related via
\bd
ds^2=dt^2+K(dx^2+dy^2+dz^2),
\ed
where $K$ is a universal constant, which may be positive, zero, or negative. Here are 
some of the reasons why $K>0$ can be excluded: (i)~Ubiquitous causal loops; reversing 
an object's motion in time is as easy as changing its direction of motion in space. (ii)~The 
nonexistence of an invariant speed---a speed that is independent of the inertial frame in 
which it is measured---rules out massless particles, long-range forces, and the possibility 
of resting the spatial part of the metric on the cyclic behavior of particles (the rates at 
which they tick).

If $K$ is not positive, causal loops are ruled out by the existence of an invariant speed. 
For $K=0$ this is infinite, and for negative $K$ it is $c=\sqrt{1/|K|}$. The problem with 
the nonrelativistic case $K=0$ is that the rates at which free particles tick cannot fix the 
spatial part of the metric. They just define a universal inertial time scale via $ds=dt$. 
Nor are light signals available for converting time units into space units. Nor do we get 
interference from free particles since $ds=dt$ implies that $\exp(ibs[{\cal C}])$ is the same for 
all paths with identical endpoints---and without interference QM is inconsistent (with the 
existence of a macroworld, which it presupposes). Yet all there is to fix the spatial part of the 
metric---for every inertial frame---is the rates at which free or freely falling particles tick. Hence 
these ought to be invariant, and this requires negative $K$ or a finite invariant speed.

The rates at which particles tick and the correlations between events in null separation 
not only underlie the metric properties of the world but also make it possible to 
influence the behavior of particles by influencing their propagators. The only way of 
influencing the probability of finding at one space-time location a scalar particle last 
``seen'' at another location, is to modify the rate at which it ticks as it travels along each 
path connecting the two locations. The number of ticks associated with a path defines a 
species-specific Finsler geometry $dS(dt,d\bfr,t,\bfr)$~\cite{Franz,Finsler}. By invoking 
again the stability of matter (Sec.~\ref{Indef}), it can be shown that there are just two 
ways of influencing the Finsler geometry that goes with a scalar particle.

The stability of matter rests on the exclusion principle~\cite{Lieb}. For this to hold, the 
ultimate constituents of matter must be indistinguishable members of one or several 
species of fermions. The necessary indistinguishability requires that all free 
particles belonging to the same species of fermions tick at the same rate. This 
guarantees the possibility of a global system of spacetime units~\cite{MW}: While there may 
be no global inertial frame, there will be local ones, and they will mesh with each other as 
described by a pseudo-Riemannian spacetime geometry. Accordingly there is a species-specific 
way of influencing the Finsler geometry associated with a scalar particle that bends geodesics 
relative to local inertial frames, and there is a universal way that bends the geodesics of the 
pseudo-Riemannian spacetime geometry. In natural units:
\bd
dS=m\sqrt{g_{\mu\nu} dx^\mu dx^\nu}+qA_\nu dx^\nu.
\ed
Here the one-form $A$ and the tensor $g$ [of type (0,2)] represent the possible effects on the 
motion of scalar particles---{\it any\/} effects, whatever the causes may 
be~\cite{Mohrhoff97,Mohrhoff99}. If the sources of these fields have no definite positions, the 
fields themselves cannot have definite values. We take this into account by summing over 
histories of $A$ and~$g$, and for consistency with the existence of a macroworld we make sure 
that a unique history is obtained in the classical limit. Obvious and well-known constraints then 
uniquely determine the terms that we need to add to~$dS$ in order to obtain a definite field 
history in this limit (except for a possible cosmological term).

A note on the quantization of particle fields (``second quantization''):
We know from Huygens' principle that a 
sum over space-time trajectories can be replaced by a wave equation. A relativistic wave 
equation has ``negative energy'' solutions corresponding to particles for which proper 
time decreases as inertial time increases~\cite{CMR}, and it conserves charge rather 
than probability. Particle numbers are therefore variables, and variables that are not 
sharply and continuously monitored by the macroworld (a.k.a. the environment) are 
fuzzy. To accommodate fuzzy particle numbers we sum over the histories of a field the 
Lagrangian of which yields the wave equation in the classical limit. This turns the field 
modes into harmonic oscillators the quanta of which represent individual particles with 
definite energies and momenta. Expanding the interaction part of $\exp(iS)$ yields a 
sum over histories in which free particles are created and/or annihilated, and by using 
the appropriate wave equation for spin-$1/2$ particles we arrive at the Feynman rules 
for QED.

I mention this to emphasize that quantum field theory (QFT) is a method of calculating 
(multi-)particle propagators. There is no discernible reason for endowing quantum fields 
with a special ontological status. Since this is nevertheless frequently done, I shall devote 
the following section to questions of ontology.

\section{\large ONTOLOGY}
\label{ONT}
Wilczek creates the impression that what Misner, Thorne, and Wheeler call ``the 
miraculous identity of particles of the same 
type''~\cite{Misneretal1973} is explained by the fact that quantum fields, rather than 
individual particles, are the primary reality: ``We understand [the identity of two 
electrons] as a consequence of the fact that both are excitations of the same underlying 
ur-stuff, the electron field. The electron field is thus the primary reality''~\cite{Wilczek}. 
Given that the self-existent and intrinsically differentiated spatiotemporal background 
of classical field theory is not a feature that can be consistently attributed to the 
quantum world, the assignment of ontological primacy to fields is a nonstarter.

For centuries philosophers have argued over the existence of intrinsically distinct 
substances. QM has settled the question for good: There are no intrinsically distinct 
substances. The concept of substance betokens {\it existence\/}; it {\it never\/} betokens 
individuality. Individuality is strictly a matter of properties.

If you think that QM is about regularities in sensory experience or about experimental 
``interventions into the course of Nature''~\cite{FuPer}, you don't need the concept of 
``substance.'' You need it if you want to think of the quantum world as a 
free-standing reality, inasmuch as it is the concept of ``substance'' that betokens 
independent existence. And you need to know how the quantum world relates to its 
substance. Since it would be absurd to substantialize a probability algorithm, 
substantiality can't be attached to a state vector or a wave function. Nor can it be 
attached to the points of a space-time manifold. Nor can the substance of the quantum 
world be decomposed into a multiplicity of intrinsically distinct substances, as we just 
saw.

If the property of being here and the property of being there are simultaneously 
possessed, how many substances does that make? The correct answer is one, for the 
substance that betokens the reality of the property of being here also betokens the reality 
of the property of being there. QM does not permit us to interpose a multiplicity of 
distinct substances between the substance that betokens existence and the multiplicity of 
possessed positions. If particles are distinct, they are so by virtue of distinguishing properties. If 
there are no distinguishing characteristics, there is multiplicity without distinctness. The 
constituents of a Bose-Einstein condensate are many without being distinct. Their multiplicity is 
not a multiplicity of substances but the property of a single substance. When particles of the 
same type possess distinct positions, there exists a multiplicity of positions, not a multiplicity of 
substances. When the particles belong to different species, the distinct positions are correlated 
with different species-specific properties, but there never exists a multiplicity of substances. 
Treating particles as a multiplicity of substances, and hence as distinct by virtue of being a 
multiplicity of substances, rather than by virtue of possessing distinguishing properties, 
inevitably leads to the wrong statistics in situations in which distinguishing properties do not 
exist, as is well known.

QM thus lends unstinting support to the constitutive idea of all monistic ontologies: Ultimately 
there is only one substance (that is, only one thing that exists by itself, rather by virtue of 
something else). As physicists we are not concerned with the intrinsic nature of this 
substance. (It arguably plays an important role in the emergence of consciousness). What is of 
interest to us is how it acquires the aspect of a spatiotemporal expanse teeming with quarks and 
leptons. In broad outline the answer is simple enough: By entering into spatial relations with 
itself, this substance acquires at one stroke the aspect of a multiplicity of spatial 
relations, which constitute space, and the aspect of a multiplicity of formless relata, 
which constitute matter. And if you allow the spatial relations to change, you've got time 
as well, for change and time imply each other. (In a timeless world nothing can change, 
and a world in which nothing changes is a world without temporal relations; such a 
world is temporally undifferentiated and therefore timeless, just as a world without 
spatial relations is spatially undifferentiated and therefore spaceless.)

The title of this letter may give the impression that the quantum world owes its nonclassical 
aspects to the nonclassical nature of space or spacetime, rather than to the nonclassical nature 
of matter. This impression rests on a false opposition. Since space---the totality of spatial 
relations that exist between particles---does not exist in the absence of particles, the 
nonclassical ``nature of space'' is one (and only one) aspect of the nonclassical nature of 
matter. Another such aspect is the radical unity of substance pointed out in this section. The 
overarching principle is {\it logical\/}. It consists in the impossibility of objectifying all the 
distinctions we make, whether they be spatial, temporal, or substantial. It finds its formal 
expression in the necessity of summing amplitudes. Whenever QM requires us to do this, it is 
because the distinctions we make between the corresponding histories have no counterparts in 
the physical world. When we sum over histories with swapped particle identities, it is because the 
particles lack identities: The distinction between {\it this\/} particle and {\it that\/} particle 
(over and above the distinction between {\it this\/} property and {\it that\/} property) is a 
distinction that nature does not make.

\section{\large A QUANTUM THEORY OF GRAVITY?}
\label{QG}
The uncertainty principle implies that at finite energies particles cannot be brought 
arbitrarily close to each other, even if it made sense to consider arbitrarily small distances. The 
fuzziness of the metric implies that arbitrarily small distances do not exist. It thus provides a 
natural high-energy cutoff. Renormalization probably only makes sense because of this natural 
cutoff~\cite{Georgi89}. It allows us to follow the scale-dependent parameters of a 
renormalizable theory down to where the ``uncertainty'' in a distance is of the same order of 
magnitude as the distance itself, and the concept of ``scale'' loses its meaning. (``Uncertainty'' 
mistranslates Heisenberg's term ``Unsch\"arfe,'' which means ``fuzziness.'') This suggests to 
me that even if a renormalizable quantum theory of gravity existed, it wouldn't make sense, 
owing to the nonexistence of another fuzzy variable providing a natural cutoff for such a theory.

Worse than ``uncertainty'' (in lieu of ``fuzziness'') is the widespread use of ``fluctuations,'' a 
term that refers to the statistical {\it consequences\/} of the fuzziness of a variable. Recall that 
the fuzziness of a variable finds expression as a statistical distribution over the possible results of 
unperformed measurements. If the measurements are performed, the fuzziness evinces itself 
counterfactually, as something that would have been had the measurements not been made 
(Sec.~2). As a rule, the indefiniteness of a variable cannot evince itself, through statistically 
distributed and hence unpredictable results, unless the results are distributed over something 
more definite. The fuzziness of the position of a material object, for instance, can evince itself 
through statistical fluctuations only to the extent that detectors with sharper positions exist. This 
is why the fuzziness of the positions of macroscopic objects cannot evince itself: Detectors with 
sharper positions do not exist.

The fuzziness of positions thus has factual consequences {\it up to a point\/}, and so has the 
fuzziness of the electromagnetic field. The indefiniteness of $A$ induces a fuzziness in the 
species-specific lengths of the ``possible'' trajectories of electrically charged particles. On all 
observationally accessible scales this fuzziness exceeds the fuzziness that is induced by the 
indefiniteness of the metric~$g$. This is why the fuzziness of $A$ can have factual consequences 
such as the Lamb shift. This effect exists not simply because a 2S electron is closer to the proton 
on average than is a 2P electron but essentially because on atomic scales ``closer'' is still 
extremely well defined. Because the 2S electron probes the electromagnetic field on a smaller 
scale than the 2P electron, it ``sees'' a fuzzier field, and thus more contributions from the 
perturbation expansion of the interaction part of $\exp(iS)$.

On the other hand, there is nothing less fuzzy than the metric. Hence the fuzziness of the metric 
has no statistical consequences in the realm of facts, for essentially the same reason that the 
fuzziness of macroscopic positions has none. Nor can there be anything comparable to the 
Lamb shift, for it is essential for this kind of effect that distinct scales exist---such as the distinct 
scales on which electrons in the aforementioned states probe the field. This suggests to me that 
the only physical consequence of the fuzziness of the metric is that it provides a natural cutoff 
for the quantum field theories of particle physics. (There is another: the very existence of GR. In 
the quantum world, everything that is not completely indefinite, and therefore nonexistent, is 
based on the interference of ``histories'' that are not objectively distinct.)

GR has been said to be an effective theory~\cite{Isham}, which suggests 
the existence of a more correct theory for scales on which the fuzziness of the metric becomes 
significant. But on such scales the very concept of ``scale'' loses its meaning. That is why a 
``quantum theory of gravity'' may be a contradiction in terms. Such a theory would make it 
possible to investigate the physics on scales that do not exist. It would allow us to study the 
physical consequences of a fuzziness lacks physical consequences, other than the 
aforesaid ones.

\section{\large ``INITIAL CONDITIONS''}
\label{Ini}
As we approach the cosmological time $t=0$, we reach a point at which the concept of 
``distance'' loses its meaning and the possibility of spatiotemporal structure ceases to 
exist. Well before we reach this point, we enter an era in which there is as yet no 
macroworld. Yet nothing happens or is the case unless it is indicated by what goes on in 
the macroworld. The properties that make up the quantum domain, including the 
properties of the universe at pre-macroscopic times, exist only to the extent that they can 
be inferred from the goings-on in the macroworld. Hence whatever happened before the 
onset of the macroworld did so only because it is indicated by something that happened 
later.

For the rest, QM allows us to make counterfactual probability assignments. As long as 
the indefiniteness of the metric does not void the concepts of ``distance'' and 
``duration,'' we can counterfactually consider regions of space that are not probed by 
any detector, and times that are not indicated by any clock. Such regions and times 
would exist if they were probed or indicated, which cannot be the case before the onset 
of the macroworld.

Finally, probabilities that are assigned to the possible results of measurements at 
pre-macroscopic times cannot be based on a ``preparation.'' QM allows us to assign 
probabilities on the basis of {\it any\/} relevant set of data, not only data pertaining to 
earlier times but also data pertaining to later times and data involving events in 
spacelike separation~\cite{Mohrhoff01}. However, where the early universe is 
concerned, there are no earlier relevant data, so probability assignments can only be 
based on data pertaining to later times. Hence the only density operator that we can 
meaningfully associate with the early universe is an advanced or ``retropared'' one---a 
density operator that ``evolves'' toward the past in the same (spurious) sense in which a 
retarded or ``prepared'' density operator ``evolves'' 
futurewards~\cite{Mohrhoff00,Mohrhoff02,AK}. The notion that the 
density operator of the early universe causally determines the later universe is therefore 
as absurd as the idea that a ``prepared'' density operator causally determines its 
``preparation.'' If we had an ensemble of early universes, we would be dealing with a 
post-selected rather than a pre-selected ensemble. We would have final 
conditions---macroscopic data relevant to probability assignments to possible events at 
pre-macroscopic times---but no initial conditions.

\end{document}